\documentclass[prb,twocolumn,showpacs, floatfix,superscriptaddress]{revtex4-1}
\usepackage{graphicx}
\usepackage{dcolumn}
\usepackage{bm}

\begin{document}

\title{Quasiparticle Nernst effect in the cuprate superconductors from the $%
d $-density wave theory of the pseudogap phase}
\author{Chuanwei Zhang}
\affiliation{Department of Physics and Astronomy, Washington State University, Pullman,
WA 99164}
\author{Sumanta Tewari}
\affiliation{Department of Physics and Astronomy, Clemson University, Clemson, SC 29634}
\author{Sudip Chakravarty}
\affiliation{Department of Physics and Astronomy, University of California, Los Angeles,
CA 90095}
\date{\today}

\begin{abstract}
We consider the Nernst effect in the underdoped regime of the cuprate high
temperature superconductors within the $d$-density wave (DDW) model of the
pseudogap phase. By a combination of analytical and numerical arguments, we
show that there is a robust low-temperature positive peak (\textit{i.e.},
maximum) in the temperature dependence of the Nernst coefficient when the
DDW state is ambipolar, \textit{i.e.}, when the broken symmetry supports the
coexistence of both electron- and hole-like quasiparticles in the excitation
spectrum, and the electron pocket dominates at the low temperatures. In
contrast, the Nernst coefficient is negative and there is no such positive
peak if the underlying state is non-ambipolar, \emph{i.e.}, when it supports
only one type of quasiparticles. More generally, in the ambipolar state, the
sign of the Nernst coefficient can be positive or negative depending on the
dominance of the electron or hole pockets, respectively, in the low
temperature thermoelectric transport. By modeling the pseudogap phase by a
doping-dependent DDW order parameter with a Fermi surface topology that
supports both hole and electron pockets, and assuming energy-independent
transport scattering times, we analyze the evolution of the Nernst effect
with doping concentration at low temperatures in the cuprate phase diagram.
Even though the chosen ambipolar DDW state with a specific Fermi surface
topology is not the only possible explanation of either the recent quantum
oscillation experiments or the recent observation of a negative Hall
coefficient at low temperatures in the underdoped cuprates, it is at least
one possible state qualitatively consistent with both of these experiments.
As such, the calculations in this paper present at least one possible
scenario for the observed enhanced Nernst signals in the underdoped cuprates.
\end{abstract}

\maketitle

\section{Introduction}

Even after two decades of intensive efforts, the normal state properties of
the cuprate superconductors in the intermediate range of hole doping, called
underdoping, are still poorly understood.\cite{Norman} At low doping, close
to the undoped antiferromagnetic phase,\cite{Halperin} the behavior of the
system is influenced by the parent Mott insulator. At doping level above
that corresponding to the maximum superconducting transition temperature ($%
T_{c}$), the mobile holes in the normal state constitute a Fermi liquid.\cite%
{Taillefer} However, at the doping range intermediate between these two
limits,
the system evinces a gap in the spectrum of unidentified origin (pseudogap)
below a temperature scale $T^{\ast }>T_{c}$. Many properties of the system
in this phase, called the pseudogap phase, are strongly influenced by the
gap, which is, similar to the superconducting gap below $T_{c}$, anisotropic
in the momentum space.
An understanding of the pseudogap, and the associated loss of the spectral
weight from the Fermi surface, is widely believed to hold the key to the
high transition temperature in the cuprates.\cite{Norman} The existence of
the gap, even in the absence of super-conduction above $T_{c}$, have led
many theorists to propose exotic non-Fermi-liquid states to be responsible
for the pseudogap in the cuprates.
However, recent quantum oscillation experiments~\cite%
{Doiron-Leyraud:2007,LeBoeuf:2007,Bangura:2008,Jaudet:2008,Yelland:2008,Sebastian:2008}
have found evidence of Fermi pockets even in the enigmatic pseudogap phase.
This has rekindled the encouraging prospect of describing this phase in
terms of a state with a broken symmetry and a reconstructed Fermi surface,%
\cite{LeBoeuf:2007,Chakravarty:2008b,Millis:2007,Podolsky:2008,
Morinari:2009, Chen:2009, Subir} treating its hole- and electron-like low
energy quasiparticles within a well-defined Fermi-liquid-like description.
Note that the Fermi arc picture, as observed in the angle resolved
photoemission (ARPES) experiments, \cite{Shen} and the Fermi pocket picture
inferred from quantum oscillation are at odds with each other, constituting
a major puzzle in the field. There have been many density wave scenarios in
which the coherence factors involved in ARPES, but not in quantum
oscillation calculations, destroy half of the pockets, giving the appearance
of a Fermi arc. \cite{Jia,Sudip-ARPES} On the other hand, ARPES has also
revealed the existence of pockets in some recent experiments. \cite%
{Meng,Yang}

One of the important unsettled questions about the pseudogap phase concerns
the low temperature Nernst effect. The Nernst effect experiments measure the
transverse electric field response of a system to a combination set-up of an
externally-imposed temperature gradient and an orthogonal magnetic field.
Early experiments \cite{Wang1,Lee2,Wang2} on the Nernst effect in the
cuprates revealed a very large signal (compared to that of a Fermi liquid)
near $T_c$, which is expected because of the presence of the large number of
mobile vortices at these temperatures. The large signal, however, appeared
to onset at a temperature far above $T_c$, leading to speculations that
there are well-defined, vortex-type, excitations even at such high
temperatures. More recent experiments \cite{Olivier} have claimed to find
two peaks in the temperature dependence of the Nernst coefficient, one
arising from the onset of a density wave order in the pseudogap phase, and
the other due to the onset of the superconducting phase. There is also
recent evidence of finding a weak peak in the Nernst signal in the pseudogap
phase whose sign is opposite to that expected from the vortex-like
excitations.\cite{chang} These recent developments, therefore, point to the
importance of the quasiparticle Nernst effect associated with an underlying
density wave state in the pseudogap regime of the cuprates. While the
experimental scenario still needs to be settled, in this paper we deduce the
full temperature and doping dependence of the quasiparticle Nernst effect
associated with the $d$-density wave state,\cite{Nayak,Chakravarty01} which
has otherwise shown encouraging consistency with the anomalous phenomenology
of the underdoped cuprates.

Since the pseudogap has a $d$-wave symmetry, one natural density wave state which could explain it is the $d_{x^{2}-y^{2}}$-density wave
state.\cite{Nayak,Chakravarty01} Indeed, much of the phenomenology of
the cuprates in the underdoped regime can be unified \cite%
{Chakravarty01,Sumanta,Chakravarty04} by making a single assumption that the
ordered DDW state is responsible for the pseudogap. The development of the
DDW order below optimal doping can lead to a consistent explanation of
numerous experimental observations including the abrupt suppression of the
superfluid density, \cite{Tewari:2001} and Hall number \cite{Tewari} below
optimal doping as well as the more recent quantum oscillation experiments.
\cite{Chakravarty:2008b} Theoretically speaking, any appropriate Hamiltonian
that leads to $d$-wave superconductivity in the underdoped regime of the
cuprates will almost certainly favor DDW order as well. \cite{Nayak:2000,
Nayak:2002} The DDW order might also have been directly observed in two
polarized neutron scattering experiments,\cite{Mook1,Mook2} even though some
other experiments failed to observe it.\cite{Stock,Fauque, footnote}
The ordered ambipolar DDW state\cite{Chakravarty:2008b} and its associated
Fermi surface topology (Fig.~\ref{fig:pockets}) are also qualitatively
consistent with the quantum oscillation experiments in the pseudogap regime.
The quantum oscillation experiments indicate that the Fermi surface in the
underdoped cuprates is made up of small reconstructed Fermi pockets, giving
rise to both hole and electron-like charge carriers (quasiparticle
ambipolarity) in the excitation spectrum. Such a feature is quite robust for
the DDW state, in which, for generic values of the band structure and gap
parameters, the low energy spectrum consists of both electron and hole-like
quasiparticles (Fig.~\ref{fig:pockets}).

We will derive the implications of the above important new ingredient in the
cuprate physics on the quasiparticle Nernst coefficient of the DDW state.
Using quasiclassical Boltzmann theory of transport, we will show that the
reconstructed Fermi surface in the DDW state and its low energy
quasiparticle ambipolarity can successfully explain the enhanced Nernst
signals as found in the experiments at temperatures much above $T_{c}$. \cite%
{Wang1,Lee2,Wang2,Olivier} Even though strong electronic interactions
present in the host material are crucial for the formation of the DDW state,
\cite{Nayak:2000, Nayak:2002} deep in the ordered state the quasiparticles
can be assumed to be non-interacting (or weakly-interacting). Therefore, we
assume that the Boltzmann theory is still applicable to calculate the
transport properties of the quasiparticles in the presence of a
well-developed DDW order parameter.

By a combination of analytical and numerical arguments, we show that a
low-temperature peak in the Nernst coefficient is very robust in the
ambipolar $d$-density wave state. In fact, the existence of the peak is
solely due to the dominance of the two types of quasiparticles (electron and
hole) at different regimes of temperatures, and is insensitive to the
microscopic details. Therefore, quasiparticle ambipolarity of the underlying
state, as indicated in the quantum oscillation experiments, is also crucial
for the low temperature peak in the Nernst coefficient. We also find that
the sign of the peak of the Nernst effect can be positive or negative
depending on the dominance of the electron or hole pockets, respectively, in
the low temperature thermoelectric transport. By modeling the pseudogap by a
suitable, doping-dependent, $d$-density wave order parameter, we analyze the
doping dependent evolution of the Nernst effect at a fixed low temperature
in a range of hole-concentrations in the underdoped regime of the cuprate
phase diagram. The quasiparticle Nernst effect has also been recently
studied \cite{Hackl} within the stripe order \cite{Kivelson} model of the
underdoped cuprates.

The paper is organized as follows: Section II introduces the commensurate
DDW state and the corresponding Hartree-Fock Hamiltonian. Section III gives
a brief description of the Nernst coefficient. Section IV is devoted to the
temperature dependence of the Nernst coefficient. We find, both numerically
and analytically, that there is a positive peak of the Nernst signal for the
ambipolar DDW state when the electron pocket dominates the transport at low
temperatures. In contrast, there is no such peak of the Nernst effect if the
underlying state is non-ambipolar. In fact, the Nernst signal from
individual electron or hole pockets are both negative, while the combination
of them can lead to a positive peak. In Section V, we discuss the doping
dependence of the Nernst coefficient. A positive peak in the Nernst signal
as a function of hole doping is also found. Finally, we summarize and
conclude in Section VI.

The main assumptions (to be explained in more detail below) we use to derive
the results of this paper are: 1) Boltzmann theory is applicable to
calculate the transport properties of the DDW quasiparticles, 2) The
transport scattering lifetimes $\tau_e, \tau_h$ (Eqs. (\ref{eq:xy}, \ref%
{eq:xx})) in the underdoped regime are constant over the Fermi surface and
are also taken to be independent of energy in a small
(temperature-dependent) interval around the Fermi energy, 3)The underlying
band structure consists of both electron and hole pockets. The third
assumption, that of quasiparticle ambipolarity, is the most crucial one for
the qualitative robustness of the temperature and doping-dependent peaks in
the DDW Nernst coefficient. Even though the existence of both types of
quasiparticles as in the ambipolar DDW state is likely not the only
explanation of either the quantum oscillation experiments or the recent
observation of negative Hall coefficients at low $T$ in the underdoped
cuprates, \cite{LeBoeuf:2007} the ambipolar DDW state (with a specific Fermi
surface topology given in Fig.~(\ref{fig:pockets})) is at least one possible
scenario consistent with both of these experiments. As such, the Nernst
calculations in this paper within the ambipolar DDW model give at least one
possible explanation of the enhanced Nernst signals in the underdoped regime
of the cuprates.

A part of the results (numerical calculation for the temperature dependence
of the Nernst coefficient) contained in this paper were published earlier.
\cite{Tewari1} In addition to giving a more complete discussion of these
previously published results, the present paper contains the following new
results: 1) An analytical explanation for the peak in the temperature
dependence of the Nernst signal, 2) A fact that even though both electron
and hole pockets can give negative Nernst effects individually, the
combination of them can yield a positive peak as a function of temperature
and 3) The full doping dependence of the DDW Nernst coefficient in the
underdoped regime of the cuprates. Specifically, we show that there is a
well-defined positive peak in the  DDW Nernst coefficient as a function of hole doping,
which is consistent with the underdoped regime in the cuprate phase diagram.

\section{Commensurate DDW state}

The commensurate DDW state \cite{Nayak} is described by an order parameter
which is a particle-hole singlet in spin space,
\begin{equation}
\left\langle \hat{c}_{\bm k+\bm Q,\alpha }^{\dagger }\hat{c}_{\bm k,\beta
}\right\rangle \propto iW_{\bm k}\,\delta _{\alpha \beta },\;W_{\bm k}=\frac{%
W_{0}}{2}(\cos k_{x}-\cos k_{y}).  \label{Order-Parameter}
\end{equation}%
Here $\hat{c}^{\dagger }$ and $\hat{c}$ are the electron creation and
annihilation operators on the square lattice of the copper atoms, $\bm %
k=(k_{x},k_{y})$ is the two-dimensional momentum, $\bm Q=(\pi ,\pi )$ is the
wave vector of the density wave, and $\alpha $ and $\beta $ are the spin
indices. For simplicity, we have taken $\hbar =1$ and the lattice constant $%
a=1$. 
In Eq.~(\ref{Order-Parameter}), $iW_{\bm k}$ is the DDW order parameter with
the $id_{x^{2}-y^{2}}$ symmetry in the momentum space. For $\bm Q=(\pi ,\pi
) $, it is purely imaginary \cite{Nayak} and gives rise to spontaneous
currents along the bonds of the square lattice.

\begin{figure}[t]
\includegraphics[width=0.65\linewidth]{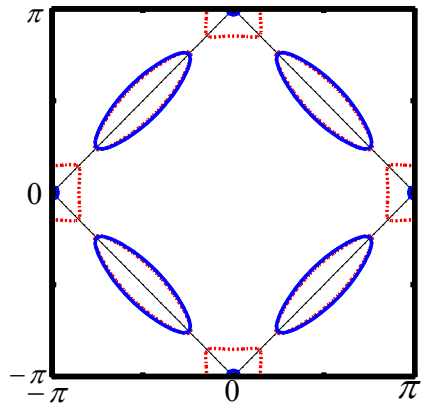}
\caption{(Color online) Electron and hole pockets in the ambipolar DDW state for two
different chemical potentials at zero temperature. Solid (blue) lines: $%
\protect\mu =-0.258$ eV ($x=10\%$); Dashed (red) lines: $\protect\mu =-0.238$
eV ($x=7\%$).}
\label{fig:pockets}
\end{figure}

The Hartree-Fock Hamiltonian describing the mean-field DDW state is given
by,
\begin{eqnarray}
&\hat{H}=\sum_{{\bm k}\in \mathrm{RBZ}}\left(
\begin{array}{cc}
\varepsilon _{\bm k}-\mu & iW_{\bm k} \\
-iW_{\bm k} & \varepsilon _{\bm k+\bm Q}-\mu%
\end{array}%
\right) ,&  \label{Hamiltonian} \\
&\varepsilon _{\bm k}=-2t(\cos k_{x}+\cos k_{y})+4t^{\prime }\cos k_{x}\cos
k_{y},&
\end{eqnarray}%
where $\varepsilon _{\bm k}$ is the band dispersion of the electrons, and $%
\mu $ is the chemical potential. The Hamiltonian density in Eq.~(\ref%
{Hamiltonian}) operates on the two-component spinor $\hat{\Psi}_{\bm k}=(%
\hat{c}_{\bm k},\hat{c}_{\bm k+\bm Q})$ defined on the reduced Brillouin
zone (RBZ) described by $k_{x}\pm k_{y}=\pm \pi $, and can be expanded over
the Pauli matrices $\hat{\bm\tau }$ and the unity matrix $\hat{I}$,
\begin{equation}
\hat{H_{\bm k}}=w_{0}(\bm k)\hat{I}+\bm w(\bm k)\cdot \hat{\bm\tau },\quad
w_{0}=\frac{\varepsilon _{\bm k}+\varepsilon _{\bm k+\bm Q}}{2}-\mu ,
\end{equation}%
where, $w_{1}=0,\quad w_{2}=-W_{\bm k},\quad w_{3}=\frac{\varepsilon _{\bm %
k}-\varepsilon _{\bm k+\bm Q}}{2}$. 
The spectrum of the Hamiltonian consists of two branches with the
eigenenergies given by,
\begin{equation}
E_{\pm }(\bm k)=w_{0}(\bm k)\pm w(\bm k),
\end{equation}%
where, $w(\bm k)=|\bm w(\bm k)|.$ For a generic set of band structure
parameters, we use $t=0.3$ eV, $t^{\prime }=0.3t$, \cite{Anderson} and $\mu $
corresponding to a non-zero hole doping, $x$, appropriate for the underdoped
regime of the cuprates, the reconstructed Fermi surface consists of two hole
pockets near the $(\pi /2,\pm \pi /2)$ points and one electron pocket near
the $(\pi ,0)$ point in the reduced Brillouin zone. The hole and the
electron pockets of the DDW state are shown in Fig.~\ref{fig:pockets} for
two different values of the chemical potential corresponding to different
values of the hole doping. The existence of both hole and electron-like
excitations in the quasiparticle spectrum generically makes this state an
ambipolar state.

\section{Nernst coefficient}

In Nernst experiments,\cite{Wang1,Lee2,Wang2,Olivier,Behnia} a temperature
gradient, $-\mathbf{\nabla }T$, is applied on the sample along the $\hat{x}$
direction. 
For such a temperature gradient, and with a magnetic field $\mathbf{B}$
along the $\hat{z}$ direction, the charge current due to quasiparticles
along $\hat{x}$ driven by $-\mathbf{\nabla }T$ produces a balancing electric
field $\mathbf{E}$. The total charge current in the presence of $\mathbf{E}$
and $-\mathbf{\ \nabla }T$ is thus given by,
\begin{equation}
J_{i}=\sigma _{ij}E_{j}+\alpha _{ij}\left( -\partial _{j}T\right) ,
\end{equation}
where $\sigma _{ij}$\ and $\alpha _{ij}$ are the electric and the
thermoelectric conductivity tensors, respectively. In the experiments, $%
\mathbf{J}$ is set to zero and the Nernst coefficient can be written as,
\begin{equation}
\nu _{N}=\frac{E_{y}}{(-\nabla T)_{x}B}=\frac{\alpha _{xy}\sigma
_{xx}-\alpha _{xx}\sigma _{xy}}{\sigma _{xx}^{2}+\sigma _{xy}^{2}},
\label{eq:1Q_xy}
\end{equation}%
where $\sigma _{ij}$ and $\alpha _{ij}$ are the electric and the
thermoelectric conductivity tensors, respectively.

For the direction of the temperature gradient as above ($T$ \emph{decreases}
in the positive $\hat{x}$ direction), and $\mathbf{B}$ in the positive $\hat{%
z}$ direction, the vortices of a superconductor produce a Nernst signal in
the positive $\hat{y}$ direction. This is because, due to entropic reasons,
the vortices flow towards the cooler end. Due to the Josephson effect, the
mobile vortices then produce a transverse electric field, $\mathbf{E}=%
\mathbf{B} \times \mathbf{v}$, which is in the positive $\hat{y}$ direction.
Note that quasiparticles in the same set up, depending on their effective
charge, would produce a transverse electric field in positive or negative $%
\hat{y}$ direction. Because of the uniqueness of the direction of the vortex
Nernst signal,\cite{Behnia} a transverse electric field in the positive $%
\hat{y}$ direction is taken as the positive Nernst signal. According to this
sign convention, the Nernst coefficient of quasiparticles is positive if it
is calculated to be so according to Eq.~(\ref{eq:1Q_xy}), where $\nu_N$ is
defined in terms of $E_y$. This modern sign convention is opposite to the
older convention sometimes also used in the literature.\cite{Nolas}
In the Nernst experiments on the high-$T_c$ cuprates, the modern sign
convention is universally used so that the vortex signal is positive by
definition.

We calculate the off-diagonal element of the conductivity tensor, $\sigma
_{xy}$, by using the solution of the semi-classical Boltzmann equation:\cite%
{Trugman}
\begin{eqnarray}
\sigma _{xy}(\mu ) &=&e^{3}B\tau _{e}^{2}\int \frac{d^{2}k}{(2\pi )^{2}}\Big[%
\frac{\partial E_{+}(k)}{\partial k_{x}}\frac{\partial E_{+}(k)}{\partial
k_{y}}\frac{\partial ^{2}E_{+}(k)}{\partial k_{x}\partial k_{y}}  \nonumber
\\
&-&\left( \frac{\partial E_{+}(k)}{\partial k_{x}}\right) ^{2}\frac{\partial
^{2}E_{+}(k)}{\partial k_{y}^{2}}\Big](-\frac{\partial f(E_{+}(k)-\mu )}{%
\partial E_{+}})  \nonumber \\
&+&(E_{+}\rightarrow E_{-};\tau _{e}\rightarrow \tau _{h}).  \label{eq:xy}
\end{eqnarray}%
Here, the momentum integrals are over the reduced Brillouin zone. In the DDW
band-structure, the electron pocket near $(\pi ,0)$ is associated with the
upper band, $E_{+}({\bm k})$. The first integral in Eq.~(\ref{eq:xy}),
therefore, embodies the contribution to $\sigma _{xy}$ due to the
electron-like quasiparticles. We have denoted the corresponding transport
scattering time as $\tau _{e}$, which, for simplicity, is taken to be
independent of the location on the electron Fermi line. The second integral
in Eq.~(\ref{eq:xy}), where $\tau _{e}$ is replaced by the scattering time
for the hole-like careers, $\tau _{h}$, calculates the contribution to $%
\sigma _{xy}$ from the hole pockets. $\tau_h$ is also taken to be constant
everywhere on the hole Fermi lines. Even though both the scattering times
can be energy-dependent, \cite{Koralek} since the Fermi surface integrals in
Eq.~(\ref{eq:xy}) and Eq.~(\ref{eq:xx}) (see below) extend only over a small
(temperature-dependent) interval around the Fermi energy, we assume $\tau_e$
and $\tau_h$ to be energy independent in our calculations.

In general, there is no obvious reason to expect $\tau _{e}=\tau _{h}$.
For a consistent interpretation of the Hall effect experiments, \cite%
{LeBoeuf:2007} it has been recently argued that the scattering times, which
are directly proportional to the career mobilities, may in fact be different
for the electron and the hole-like charge carriers. Since at low
temperatures the Hall coefficient is negative, Ref.~[%
\onlinecite{LeBoeuf:2007}] argues that, at least at low $T$, $\tau _{e}>\tau
_{h}$. With the above definition of the parameters, the diagonal element of
the conductivity tensor is given by,\cite{Trugman}
\begin{eqnarray}
\sigma _{xx}(\mu ) &=&e^{2}\tau _{e}\int \frac{d^{2}k}{(2\pi )^{2}}\left(
\frac{\partial E_{+}(k)}{\partial k_{x}}\right) ^{2}(-\frac{\partial
f(E_{+}(k)-\mu )}{\partial E_{+}})  \nonumber \\
&+&(E_{+}\rightarrow E_{-};\tau _{e}\rightarrow \tau _{h}).  \label{eq:xx}
\end{eqnarray}

From the solution of the Boltzmann equation at low $T$, the thermoelectric
tensor $\alpha _{ij}$ is related to the conductivity tensor $\sigma _{ij}$
by the Mott relation:\cite{Marder}
\begin{equation}
\alpha _{ij}=-\frac{\pi ^{2}}{3}\,\frac{k_{B}^{2}T}{e}\,\frac{\partial
\sigma _{ij}}{\partial \mu }.  \label{eq:1alpha}
\end{equation}%
Here $e>0$ is the absolute magnitude of the charge of an electron. Using
Eq.~(\ref{eq:1Q_xy}), the formula for the Nernst coefficient reduces to,
\cite{Oganesyan}
\begin{equation}
\nu _{N}=-\frac{\pi ^{2}}{3}\,\frac{k_{B}^{2}T}{eB}\,\frac{\partial \Theta
_{H}}{\partial \mu }  \label{eq:1nernst1}
\end{equation}%
Here,
\begin{equation}
\Theta _{H}=\tan ^{-1}(\frac{\sigma _{xy}}{\sigma _{xx}}).
\end{equation}%
Using Eqs.~(\ref{eq:xy},\ref{eq:xx},\ref{eq:1nernst1}) and with reasonable
phenomenological assumptions about the temperature dependence of the
scattering times and the DDW order parameter, we can now calculate $\nu _{N}$
as a function of $T$ in the ambipolar DDW state. Using a phenomenological
ansatz for the doping dependence of the DDW order parameter, and computing
the chemical potential self-consistently, we can also use the same equations
to calculate the doping dependence of $\nu _{N}$. This way we can evaluate
the evolution of the Nernst coefficient in the cuprate phase diagram within
the ambipolar DDW model.

It is important to emphasize that the negative Hall coefficient at low $T$,
as seen in the presence of strong magnetic fields in Ref.~[%
\onlinecite{LeBoeuf:2007}], does not automatically imply the presence of the
electron pockets at low enough magnetic fields. However, conversely, the
existence of the electron pockets in the band-structure is at least one
possible scenario consistent with the negative Hall coefficient. In
addition, the existence of the electron pockets is also qualitatively
consistent with the observed frequencies in the recent quantum oscillation
experiments. Our calculated enhanced Nernst coefficients are for the
ambipolar DDW state which has both electron- and hole-like quasiparticles in
the excitation spectrum. As such, the results of this paper give at least
one possible explanation of the observed enhanced Nernst signals in the
underdoped cuprates.

\section{Temperature dependence of the Nernst coefficient}

\subsection{Phenomenological temperature dependence of the parameters}

In the first step in the evaluation of the temperature dependence of the
Nernst coefficient, we have to make suitable assumptions for the behavior of
the scattering times $\tau _{e}$ and $\tau _{h}$ with temperature. An
important hint regarding this can be obtained from the recent Hall effect
experiments in Ref.~\onlinecite{LeBoeuf:2007}. In these experiments, the
normal state Hall coefficient,
\begin{equation}
R_{H}=\frac{\sigma _{xy}}{B(\sigma _{xx})^{2}},
\end{equation}%
has been measured as a function of $T$ in three different samples of
underdoped YBCO.
In all three samples, $R_{H}$ is large and positive above $T^{\ast }$, which
is consistent with the systems being moderately hole doped. $R_{H}$,
however, shows a sharp decline below $T^{\ast }$, and subsequently changes
its sign from positive (hole-dominated) to negative (electron-dominated) at
a crossover temperature $T_{0}<T^{\ast }$.
This anomalous $T$-dependence of the Hall coefficient can be understood
naturally if the state in question below $T^{\ast }$ is inherently
ambipolar, and the mobilities of the oppositely charged quasiparticles are
assumed to be unequal and changing with temperature.

With Eqs.~(\ref{eq:xy},\ref{eq:xx}), we can calculate the contributions of
the electron and hole pockets of the DDW state to the normal state Hall
coefficient. The magnitudes of the individual contributions depend on the
size and curvature of the respective pockets, but in our calculations the
sign of the contribution is positive for the hole-like quasiparticles and
negative for the electron-like quasiparticles. For $\tau _{e}=\tau _{h}$, in
which case the formula for $R_{H}$ is independent of the scattering time,
and for a generic set of parameters ($t,t^{\prime },W_{0}$) consistent with
the quantum oscillation experiments in YBCO, \cite{Chakravarty:2008b,Jia}
the size and curvature of the hole pockets are much bigger than those of the
electron pocket. This implies that, for $\tau_{e}\sim\tau_{h}$, the sign of
the overall $R_{H}$ is positive. \cite{Tewari} We have checked that
reasonable modifications of the band structure parameters in the cuprate
phase diagram cannot change this result. Therefore, within the Boltzmann
theory of transport, the only source of the strong $T$-dependence of $R_H$,
as observed in the experiments, must come from the unequal temperature
dependence of $\tau_e$ and $\tau_h$. 
If at high temperatures ($T>T_{0}$) $\tau _{e}\sim \tau _{h}$, the Hall
coefficient is positive. On the other hand, if $\tau _{e}> \tau _{h}$ for $%
T<T_{0}$, $R_{H}$ can become negative at low $T$. Note that a higher
mobility of the electron-like quasiparticles at low $T$ is also consistent
with the frequency observed in the quantum oscillation experiments. \cite%
{LeBoeuf:2007,Chakravarty:2008b} An independent, microscopic, justification
of the higher lifetime of the electron-like quasiparticles at low $T$ and in
the presence of a magnetic field also follows by considering the scattering
of both types of quasiparticles by vortices at low temperatures. \cite%
{Stephen} Because of the difference of the effective masses between the DDW
quasiparticles near the antinodal and the nodal regions in the Brillouin
zone, the electron lifetime due to vortex scattering can be significantly
higher than the hole lifetime at low $T$ in the presence of a magnetic
field. \cite{Pallab}

On the above grounds, we choose the minimal $T$-dependence of the scattering
times:
\begin{equation}
\hbar\tau_{e}^{-1}=A_{e}+B_{e}k_{B}T  \label{taue}
\end{equation}
and
\begin{equation}
\hbar \tau _{h}^{-1}=A_{h}+B_{h}k_{B}T  \label{tauh}
\end{equation}
Even though such a linear $T$-dependence of the scattering times is
nominally consistent with the linear $T$-dependence of the resistivity in a
regime close to the optimal doping in the cuprate phase diagram, we
emphasize that our motivation for the assumptions about $\tau _{e}$ and $%
\tau _{h}$ is strictly phenomenological. With these choices, the calculated $%
R_{H}=\frac{\sigma _{xy}}{B\sigma _{xx}^{2}}$ for the ambipolar DDW state as
a function of $T$ with an assumed mean field $T$-dependence of the DDW order
parameter,
\begin{equation}
W_{0}(T)=W_{0}\sqrt{{1-\frac{T}{T^{\ast }}}}
\end{equation}
($T^{\ast }\sim 110$ K), qualitatively agrees with the recent experiments.
\cite{LeBoeuf:2007} We estimate the values of the temperature independent
parameters $A_{e}, B_{e}, A_{h}, B_{h}$ in Eq.~(\ref{taue}) and Eq.~(\ref%
{tauh}) from the qualitative agreement of the Hall effect experiments in the
underdoped cuprates. We do this by setting the total $R_H$ to zero at $%
T=T_0=30$ K. This provides one equation relating the four unknown constants.
Then we assume that at high temperatures $\tau_e$ approximately equals $%
\tau_h$, which stipulates $B_e \sim B_h$. We take them both to be equal to $%
1 $ for simplicity. Therefore, we are left with two unknowns $A_e$ and $A_h$
and only one equation relating them, which leaves some residual freedom in
choosing the values of these constants. However, we have checked that our
conclusions for the behavior of $\nu $ with $T$ are robust to any reasonable
variation of the $T$-dependence of $\tau _{e},\tau _{h}$ and $W_{0}(T)$ as
long as they satisfy the experimental constraints set by the temperature
dependence of $R_H$.

\begin{figure}[t]
\includegraphics[width=0.65\linewidth]{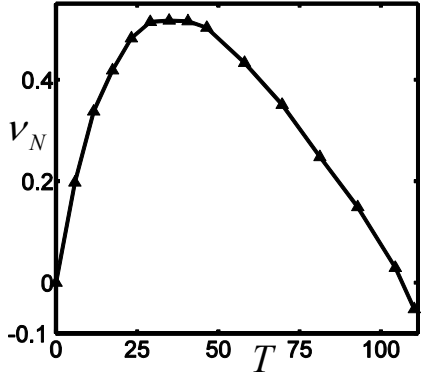}
\caption{Plot of the Nernst coefficient, $\protect\nu _{N}$, versus
temperature $T$ for the ambipolar DDW state. The Nernst coefficient in V K$%
^{-1}$T$^{-1}$ can be derived by multiplying the dimensionless $\protect\nu %
_{N}$ in the figure with the quantity $2\protect\pi ^{2}k_{B}a^{2}/3\hbar
\approx 138$ nV/KT. Here the lattice constant is taken as $a \sim 0.4$ nm,
and the factor 2 is used to account for the contributions from the two spin
components. The peak value is $\sim 70$ nV/KT at a temperature $T<T^{\ast }$%
. The unit for $T$ along the horizontal axis is $K$. The hole doping $x=10\%$%
. The sign of $\protect\nu _{N}$ near its peak is positive. As the
superconducting $T_{c}$ (not shown here) is approached, the normal state $%
\protect\nu _{N}$ as shown here will be cut-off by the large Nernst signal
of the mobile vortices associated with the superconductor. The parameters
used in the plot are $T^{\ast }=110$ K, $A_{e}/k_{B}=58$ K, $A_{h}/k_{B}=$
253 K, $t=0.3$ eV, $t^{\prime }=0.3t$, $B_{e}=B_{h}=1$, $W_{0}=0.1$ eV, $%
\protect\mu =-0.258$ eV.}
\label{fig:NernstT1}
\end{figure}

We now use Eqs.~(\ref{eq:xy}, \ref{eq:xx}, \ref{eq:1nernst1}) to calculate $%
\nu _{N}$ as a function of $T$ for a specific value of the hole doping $%
x=10\% $. Using a mean field $T$-dependence of $W_{0}(T)$ and the
phenomenological form of $\tau _{e}(T)$ and $\tau _{h}(T)$ above, we plot in
Fig.~(\ref{fig:NernstT1}) the calculated $\nu _{N}$ in the ambipolar DDW
state as a function of $T$ for $x=10\%$. It is clear from Fig.~(\ref%
{fig:NernstT1}) that the Nernst coefficient has a pronounced low temperature
peak which, as we argue below, is a direct manifestation of the
quasiparticle ambipolarity of the DDW state.

\subsection{Sign and temperature dependence of Nernst signal from individual
hole and electron pockets}

To elucidate the importance of the quasiparticle ambipolarity in the
temperature dependence of $\nu _{N}$, let us first consider the Nernst
effect due to the quasiparticles associated with a hole pocket. In the
presence of only hole pockets in the excitation spectrum, we can write the
Nernst coefficient as
\begin{equation}
\nu _{N}^{h}=-C\tau _{h}T\frac{\partial }{\partial \mu }\bar{\Theta}^{h},
\label{nu1}
\end{equation}%
where we take
\begin{equation}
\bar{\Theta}^{h}\approx \frac{\bar{\sigma}_{xy}^{h}}{\bar{\sigma}_{xx}^{h}},%
\text{ \ \ (}\bar{\sigma}_{xy}^{h}\ll \bar{\sigma}_{xx}^{h}\text{)}
\end{equation}%
\begin{equation}
\bar{\sigma}_{xy}^{h}=\sigma _{xy}^{h}/\tau _{h}^{2},\text{ \ }\bar{\sigma}%
_{xx}^{h}=\sigma _{xx}^{h}/\tau _{h}
\end{equation}
and $C$ is a numerical constant, $C=\frac{\pi ^{2}}{3}\frac{k_{B}^{2}}{eB}$.
$\bar{\sigma}_{xy}^{h}$ and $\bar{\sigma}_{xx}^{h}$ (superscript \emph{h}
indicates the contribution from the hole-like quasiparticles) depend only on
the hole Fermi surface integrals in Eq.~(\ref{eq:xy}) and Eq.~(\ref{eq:xx}),
respectively. We have rewritten the expression for $\nu _{N}$ (Eq.~(\ref%
{eq:1nernst1})) in Eq.~(\ref{nu1}) so that the manipulation of the explicit $%
T$-dependence of the scattering time becomes easier. Similarly, when there
are only electron-like quasiparticles in the system, the Nernst coefficient
can be written as,
\begin{equation}
\nu _{N}^{e}=-C\tau _{e}T\frac{\partial }{\partial \mu }\bar{\Theta}^{e},
\end{equation}%
where $\bar{\Theta}^{e}\approx \frac{\bar{\sigma}_{xy}^{e}}{\bar{\sigma}%
_{xx}^{e}}$, $\bar{\sigma}_{xy}^{e}=\sigma _{xy}^{e}/\tau _{e}^{2}$, $\bar{%
\sigma}_{xx}^{e}=\sigma _{xx}^{e}/\tau _{e},$ and the superscript \emph{e}
indicates electron-like quasiparticles.

\begin{figure}[t]
\includegraphics[width=0.75\linewidth]{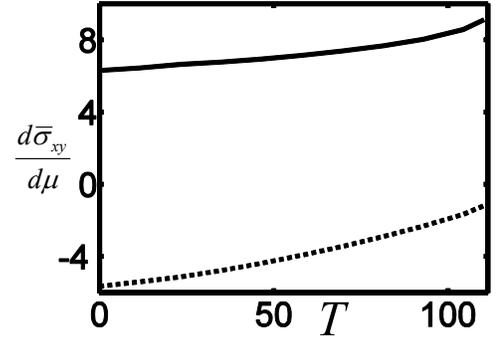}
\caption{Plot of $\frac{d\bar{\protect\sigma}_{xy}}{d\protect\mu }$ versus
temperature for the electron and hole pockets using Eqs. (8,19). The unit of
$T$ is $K$, and the unit for $\frac{d\bar{\protect\sigma}_{xy}}{d\protect\mu
}$ is unimportant for the purpose of this illustration. $x=10\%$. Dashed
line: electron; Solid line: hole. }
\label{fig:dsdu}
\end{figure}

\begin{figure}[b]
\includegraphics[width=0.99\linewidth]{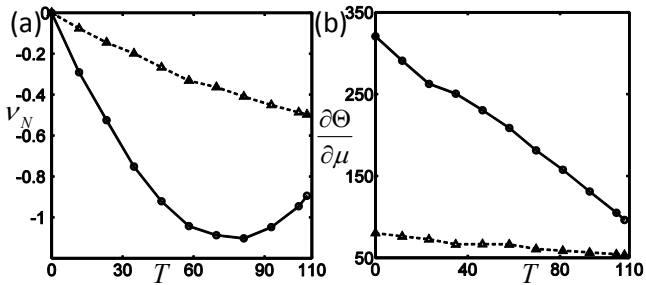}
\caption{(a) Plot of the Nernst coefficient, $\protect\nu _{N}$, versus
temperature for the electron (solid line) and hole (dashed line) pockets
separately. The parameters used in the plot are same as those in Fig.
\protect\ref{fig:NernstT1}. Nernst coefficient in V K$^{-1}$T$^{-1}$ can be
derived by multiplying the dimensionless $\protect\nu _{N}$ in the figure
with the factor $2\protect\pi ^{2}k_{B}a^{2}/3\hbar \approx 138$ nV/KT. The
temperature in the horizontal axis is in $K$. The hole doping $x=10\%$. (b)
Plots of $\frac{\partial \Theta }{\partial \protect\mu }$ versus temperature
for the electron and hole pockets separately. The plots in (a) are obtained
by multiplying these values by $-k_{B}T$ in eV. }
\label{fig:nersep}
\end{figure}

In the following analysis, we will frequently need the sign of the
quantities, $\frac{d\bar{\sigma}_{xy}^{h}}{d\mu },\frac{d\bar{\sigma}%
_{xx}^{h}}{d\mu }$, and their counterparts for the electron pockets. These
can be deduced by noting the changes in the shapes of the hole and the
electron pockets with increasing $\mu $ ($\mu $ becoming less negative). To
do this, we recall that the magnitude of $\bar{\sigma}_{xy}$ increases with
the curvature of the relevant pocket, and the magnitude of $\bar{\sigma}%
_{xx} $ increases with its area. In Fig.~\ref{fig:pockets}, we plot the
electron and hole pockets for two different values of the chemical
potential. As we can see, as $\mu $ increases,
the hole pockets become more elliptical (i.e., the curvature rises and the
circumference decreases), thus $\frac{d\bar{\sigma}_{xy}^{h}}{d\mu }>0$ and $%
\frac{d\bar{\sigma}_{xx}^{h}}{d\mu }<0$.
On the other hand, for the electron pocket, the size of the pocket increases
with increasing $\mu $, leading to $\frac{d\bar{\sigma}_{xx}^{e}}{d\mu }>0$.
It is not clear, however, how the curvature of the electron pocket varies
with $\mu $. In Fig.~(\ref{fig:dsdu}), we plot the $\frac{d\bar{\sigma}%
_{xy}^{e}}{d\mu }$ and $\frac{d\bar{\sigma}_{xy}^{h}}{d\mu }$ for $x=10\%$.
We clearly see that $\frac{d\bar{\sigma}_{xy}^{e}}{d\mu }<0$ \ and $\frac{d%
\bar{\sigma}_{xy}^{h}}{d\mu }>0$. For the hole pockets, the above behaviors
lead to
\begin{equation}
\frac{\partial \bar{\Theta}^{h}}{\partial \mu }=\frac{1}{(\bar{\sigma}%
_{xx}^{h})^{2}}(\frac{d\bar{\sigma}_{xy}^{h}}{d\mu }\bar{\sigma}_{xx}^{h}-%
\frac{d\bar{\sigma}_{xx}^{h}}{d\mu }\bar{\sigma}_{xy}^{h})>0.
\end{equation}%
This implies that the hole pockets lead to a negative Nernst signal, as seen
from Fig. \ref{fig:nersep}. The analysis of the sign of the Nernst
coefficient from the electron pocket is not so straightforward, and will be
discussed later.

Taking the temperature derivative of $\nu _{N}^{h}$ in Eq.~(\ref{nu1}), we
get,
\begin{eqnarray}
C^{-1}\frac{\partial \nu _{N}^{h}}{\partial T} &=&-\tau _{h}\frac{\partial
\bar{\Theta}^{h}}{\partial \mu }-T\frac{\partial \tau _{h}}{\partial T}\frac{%
\partial \bar{\Theta}^{h}}{\partial \mu }-T\tau _{h}\frac{\partial }{%
\partial T}\frac{\partial \bar{\Theta}^{h}}{\partial \mu }\   \nonumber \\
&=&-\frac{1}{T+A_{h}}\left( 1-\frac{T}{T+A_{h}}\right) \frac{\partial \bar{%
\Theta}^{h}}{\partial \mu }  \nonumber \\
&&-T\tau _{h}\frac{\partial }{\partial T}\frac{\partial \bar{\Theta}^{h}}{%
\partial \mu },  \label{eq:derivative}
\end{eqnarray}%
where we take $B_{h}=1$. In Eq.~(\ref{eq:derivative}), the first term on the
right hand side is the result of the explicit $T$-dependence of $\nu _{N}$
via the scattering time and the explicit factor of $T$. Since $\frac{%
\partial \bar{\Theta}^{h}}{\partial \mu }$ is positive,
this term is strictly negative. The second term in Eq.~(\ref{eq:derivative})
depends on the implicit $T$-dependence of $\nu _{N}^{h}$ via the $T$%
-dependence of $W_{0}$. To calculate this term, we write,
\begin{equation}
\frac{\partial }{\partial T}\frac{\partial \bar{\Theta}^{h}}{\partial \mu }=%
\frac{\partial }{\partial \mu }\frac{\partial \bar{\Theta}^{h}}{\partial T}=%
\frac{\partial }{\partial \mu }\Big(\frac{\partial \bar{\Theta}^{h}}{%
\partial W_{0}}\frac{-1}{2\left( T^{\ast }-T\right) ^{1/2}}\Big),
\label{eq:sign0}
\end{equation}%
where we have used the mean field ansatz for the $T$-dependence of the
amplitude of the DDW order parameter. To calculate the right hand side,
noting that $T$ is independent of $\mu $, we only need to calculate,
\begin{eqnarray}
\frac{\partial }{\partial \mu }\frac{\partial \bar{\Theta}^{h}}{\partial
W_{0}} &=&\frac{\partial }{\partial \mu }\frac{1}{\bar{\sigma}_{xx}^{h}}%
\left( \frac{d\bar{\sigma}_{xy}^{h}}{dW_{0}}-\bar{\Theta}^{h}\frac{d\bar{%
\sigma}_{xx}^{h}}{dW_{0}}\right)  \nonumber \\
&\approx &-\frac{1}{(\bar{\sigma}_{xx}^{h})^{2}}\frac{\partial \bar{\sigma}%
_{xx}^{h}}{\partial \mu }\left( \frac{d\bar{\sigma}_{xy}^{h}}{dW_{0}}-\bar{%
\Theta}^{h}\frac{d\bar{\sigma}_{xx}^{h}}{dW_{0}}\right)  \nonumber \\
&&-\frac{1}{\bar{\sigma}_{xx}^{h}}\frac{\partial \bar{\Theta}^{h}}{\partial
\mu }\frac{d\bar{\sigma}_{xx}^{h}}{dW_{0}}.  \label{eq:sign}
\end{eqnarray}

In the derivation of the above equation, we neglect the small terms $\frac{%
\partial }{\partial \mu }\frac{d\bar{\sigma}_{xy}^{h}}{dW_{0}}$ and $\frac{%
\partial }{\partial \mu }\frac{d\bar{\sigma}_{xx}^{h}}{dW_{0}}$ (see Eq. (%
\ref{sig}) below for justification). It is clear that we first need the
leading $W_{0}$-dependence of the Fermi surface integrals $\bar{\sigma}%
_{xy}^{h}$ and $\bar{\sigma}_{xx}^{h}$. We find that the leading $W_{0}$
dependence of these integrals is linear. This linear dependence arises from
the integration region around the so-called \emph{hot spots}, points on the
Fermi surface which also fall on the surface of the RBZ. On the surface of
the RBZ ($k_{x}\pm k_{y}=\pm \pi $), the dominant part of the band
structure, $-2t(\cos k_{x}+\cos k_{y})=0$. The linear $W_{0}$ dependence of $%
\bar{\sigma}_{xy}^{h}$ and $\bar{\sigma}_{xx}^{h}$ comes from the region
around the hot spots which satisfies
\[
t|\cos k_{x}+\cos k_{y}|<\frac{W_{0}}{4}|\cos k_{x}-\cos k_{y}|,
\]%
which has a size $\sim tW_{0}$. Since the relevant integration region is
itself ${\mathcal{O}}(W_{0})$, by expanding the integrand in Eq.~(\ref{eq:xy}%
) to zeroth order in $W_{0}$, we get the leading $W_{0}$ dependence of $\bar{%
\sigma}_{xy}^{h}$ as,
\begin{equation}
\bar{\sigma}_{xy}^{h}\sim 64tW_{0}{t^{\prime }}^{3}\sin ^{2}k_{x}\cos
k_{x}\cos k_{y}.  \label{sig}
\end{equation}%
The above is negative since $\cos k_{x}=-\cos k_{y}$ on the surface of RBZ.
Therefore, it follows that, $\frac{d\bar{\sigma}_{xy}^{h}}{dW_{0}}<0$. From
similar manipulations, it is straightforward to show that $\frac{d\bar{\sigma%
}_{xx}^{h}}{dW_{0}}>0$. We also note that, in Eq.~(\ref{eq:sign}), $\bar{%
\Theta}^{h}$ itself is positive for the hole pockets ($\bar{\Theta}^{h}$ has
the same sign as the Fermi surface integral for the Hall conductivity, $\bar{%
\sigma}_{xy}^{h}$, which is, of course, positive for the hole pockets).
Furthermore, noting that $\frac{\partial \bar{\sigma}_{xx}^{h}}{\partial \mu
}<0$ and $\frac{\partial \bar{\Theta}^{h}}{\partial \mu }>0$ for the hole
pocket, we infer from Eq.~(\ref{eq:sign}) that $\frac{\partial }{\partial
\mu }\frac{\partial \bar{\Theta}^{h}}{\partial W_{0}}<0$. Therefore, from
Eq.~(\ref{eq:sign0}),
\begin{equation}
\frac{\partial }{\partial T}\frac{\partial \bar{\Theta}^{h}}{\partial \mu }=%
\frac{\partial }{\partial \mu }\frac{\partial \bar{\Theta}^{h}}{\partial T}>0
\end{equation}
for $T<T^{\ast }$. Finally, using Eq.~(\ref{eq:derivative}), we conclude
that $\frac{\partial \nu _{N}^{h}}{\partial T}$ is negative definite for the
hole pockets. This implies that, for only hole-type quasiparticles in the
DDW state, the temperature-derivative of the Nernst coefficient can never be
zero: there is no low temperature peak of $\nu _{N}^{h}(T)$. This analytical
result has been confirmed by the numerical results, as seen in Fig. \ref%
{fig:nersep}.

In contrast to $\nu _{N}^{h}$, the sign of $\nu _{N}^{e}$ is hard to
determine analytically. The reason is that, in
\[
\frac{\partial \bar{\Theta}^{e}}{\partial \mu }=\frac{1}{(\bar{\sigma}%
_{xx}^{e})^{2}}(\frac{d\bar{\sigma}_{xy}^{e}}{d\mu }\bar{\sigma}_{xx}^{e}-%
\frac{d\bar{\sigma}_{xx}^{e}}{d\mu }\bar{\sigma}_{xy}^{e}),
\]
the terms $\frac{d\bar{\sigma}_{xy}^{e}}{d\mu }\bar{\sigma}_{xx}^{e}(<0)$
and $-\frac{d\bar{\sigma}_{xx}^{e}}{d\mu }\bar{\sigma}_{xy}^{e}(>0)$ have
opposite signs, and the sign of $\nu _{N}^{e}$ should be determined by the
relative magnitudes of these two terms. For the band structure parameters
used here, we find that $\nu _{N}^{e}$ is negative, as seen in Fig. \ref%
{fig:nersep}. This implies that, for the electron pocket, the second term in
$\frac{\partial \bar{\Theta}^{e}}{\partial \mu }$ 
dominates over the first one,
leading to a positive $\frac{\partial \bar{\Theta}^{e}}{\partial \mu }$.

\subsection{Temperature dependence of the Nernst coefficient from ambipolar
DDW state}

In view of the above analysis, a natural question is then why two
individually negative contributions from the electron and the hole pockets
`add up' to a positive Nernst signal at low temperatures when the two types
of pockets coexist. The underlying reason can be most clearly expressed by
writing down the formula for $\frac{\partial \Theta ^{t}}{\partial \mu }$,
where the superscript `t' now represents the total Nernst effect as given by
multiplying $\frac{\partial \Theta ^{t}}{\partial \mu }$ by $-CT$,
\begin{equation}
\frac{\partial \Theta ^{t}}{\partial \mu }=\frac{1}{({\sigma }_{xx}^{t})^{2}}%
(\frac{d{\sigma }_{xy}^{t}}{d\mu }{\sigma }_{xx}^{t}-\frac{d{\sigma }%
_{xx}^{t}}{d\mu }{\sigma }_{xy}^{t})  \label{eq:reason}
\end{equation}%
Here, $\sigma _{xx}^{t}=\sigma _{xx}^{h}+\sigma _{xx}^{e}$ and $\sigma
_{xy}^{t}=\sigma _{xy}^{h}+\sigma _{xy}^{e}$. Because of the ambipolar
spectrum, the second term in Eq.~(\ref{eq:reason}) is much smaller than the
first one (since the total Hall conductivity, $\sigma _{xy}^{t}$, is small),
and the sign of $\frac{\partial \Theta ^{t}}{\partial \mu }$ is entirely
determined by the first term. 
It follows that if the contribution from the electron pocket dominates over
that from the hole pocket, then $\frac{\partial \Theta ^{t}}{\partial \mu }$
is negative, since the first term in Eq.~(\ref{eq:reason}) is negative for
the electron pocket. On the other hand, if the contribution from the hole
pocket is greater than that from the electron pocket, then $\frac{\partial
\Theta ^{t}}{\partial \mu }$ is positive because $\frac{d\sigma_{xy}^h}{d\mu}
$ is positive for the hole pocket, see Fig.~\ref{fig:dsdu}. In our
calculations, the former is the situation at low $T$, and $\nu _{N}$ is
positive at low temperatures.
At high temperatures, the contribution from the hole pockets dominates
transport because of their larger size and the first term in Eq.~(\ref%
{eq:reason}), and consequently $\frac{\partial \Theta ^{t}}{\partial \mu }$,
becomes positive, leading to a negative Nernst signal. Therefore, $\nu _{N}$%
, which is zero at $T=0$, first increases at low temperatures and then
decreases at high temperatures, yielding a positive peak if there are both
electron- and hole-type quasiparticles present in the spectrum at the same
time. 

In the context of the cuprates, at low temperatures, the electron pocket
dominates via $\tau _{e}>>\tau _{h}$, and $\frac{\partial \nu_{N}}{\partial T%
}>0$. On the other hand, at high temperatures, the hole pockets dominate
when $\tau _{e}\sim \tau _{h}$.
In this case, we have $\frac{\partial \nu _{N}}{\partial T}<0$.
In practice, determining an analytical expression for the peak temperature
by solving the implicit equation, $\frac{\partial \nu_{N}}{ \partial T}=0$,
is not very illuminating, since it depends on many parameters. However, we
have conclusively shown here that the existence of a low temperature
positive peak in the Nernst coefficient is a robust consequence of
quasiparticle ambipolarity, and, therefore, is independent of any
assumptions about the microscopic parameters. Furthermore, the sign of the
peak value of $\nu _{N}(T)$ depends on the relative dominance of the
electron and the hole pockets at different regimes of $T$. For example, if
the hole (electron) pockets were dominant at low (high) temperatures, the
peak value of $\nu _{N}(T)$ would be negative. However, in the
experimentally relevant case where the electron pocket is more dominant at
low $T$ (so that the zero temperature Hall coefficient is negative), the
peak is on the positive side.

\section{Doping dependence of the Nernst coefficient}

To calculate the Nernst coefficient as a function of doping in the
underdoped regime, we have to start by assuming a phenomenological doping
dependence of the DDW order parameter $W_{0}$. Using the values of the
doping-dependent amplitude of the DDW order parameter and the set of
parameters $t,t^{\prime },x$, we can calculate the chemical potential $\mu $
as a function of $x$. This way, the Nernst coefficient can be calculated as
a function of the doping in the pseudogap phase.
\begin{figure}[tbp]
\includegraphics[width=0.65\linewidth]{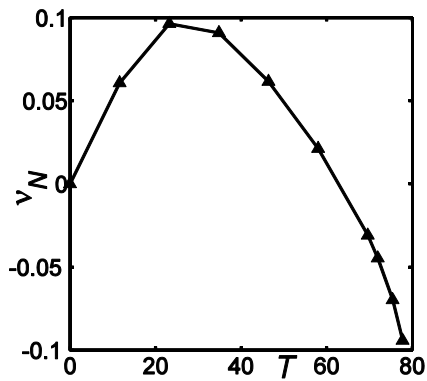}
\caption{The Nernst coefficient as a function of temperature for $x=7\%$.
Nernst coefficient in V K$^{-1}$T$^{-1}$ can be derived by multiplying $%
\protect\nu _{N}$ with the factor $2\protect\pi ^{2}k_{B}a^{2}/3\hbar
\approx 138$ nV/KT. The temperature is in $K$. The corresponding $\protect%
\mu =-0.238$ eV, $T^{\ast }=80$ K. We see that the Nernst signal is negative
in a broader temperature regime than that for $x=10\%$.}
\label{fig:NernstT2}
\end{figure}
In the absence of a concrete theoretical result for the doping dependence of
the DDW order parameter, we assume the mean-field doping dependence,
\begin{equation}
W_{0}(x,T)=W_{0}(x_{0},T)\left\{
\begin{array}{c}
\sqrt{\frac{x_{\max }-x}{x_{\max }-x_{0}}}\text{ \ \ \ if }x\geq x_{0} \\
\sqrt{\frac{x-x_{\min }}{x_{0}-x_{\min }}}\text{ \ \ \ \ if }x<x_{0}%
\end{array}%
\right.
\end{equation}%
where $x_{0}=10\%$ is the doping percentage that yields the maximum DDW
order. This kind of a doping dependence is physically motivated, since we
expect the DDW order to gradually weaken for both high and low values of $x$%
. We choose $x_{\min }=4\%$ and $x_{\max }=17\%$ as the minimum and the
maximum doping where the DDW order may exist. We use the value of $x$ and
the set of parameters $t,t^{\prime },W_{0}(x,T=0)$ to self-consistently
calculate the value of $\mu $, which determines the size and curvature of
the hole and the electron pockets.
\begin{figure}[t]
\includegraphics[width=0.65\linewidth]{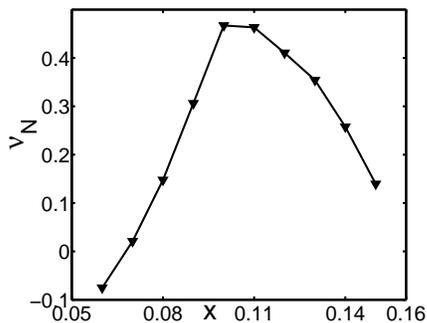}
\caption{Plot of the Nernst coefficient, $\protect\nu _{N}$, at a fixed low
temperature, $T=58$ K, versus the hole doping $x$ for the ambipolar DDW
state. Nernst coefficient in V K$^{-1}$T$^{-1}$ can be derived by
multiplying $\protect\nu _{N}$ with the factor $2\protect\pi %
^{2}k_{B}a^{2}/3\hbar \approx 138$ nV/KT. The temperature is in $K$. Nernst
coefficient shows a pronounced peak at a hole doping $x=10\%$, where the DDW
order parameter is assumed to be the largest. The sign of $\protect\nu _{N}$
near its peak is positive. }
\label{fig:Nernstx}
\end{figure}


To illustrate the behavior of $\nu _{N}$ with underdoping, we plot in Fig.~(%
\ref{fig:NernstT2}) the temperature dependence of $\nu _{N}$ for $x=7\%$. It
is clear that 
the Nernst signal remains negative in a wider range of temperatures at this
value of hole-doping than that at $x=10\%$. However, there is still a small
low temperature peak in the positive side because of the larger mobility of
the electron pocket at low temperatures. Since the Nernst effect at
temperatures close to and below the superconducting $T_{c}$ is almost
entirely dominated by the vortex Nernst signal, the low temperature positive
peak due to the DDW quasiparticles may not be visible in the experiments. In
this case, the DDW quasiparticle Nernst effect may appear negative above the
superconducting $T_{c}$.

In Fig.~(\ref{fig:Nernstx}), we plot the Nernst coefficient at a fixed low
temperature as a function of hole doping $x$. To construct this plot, we
have taken the transition temperature of the ordered DDW state to scale with
the value of the zero temperature order parameter, as would be expected from
the mean field theory. 
This implies that $T^{\ast }\left( x\right) $ has been determined via
\begin{equation}
T^{\ast }\left( x\right) =\frac{W_{0}\left( x,T=0\right) }{W_{0}\left(
x_{0},T=0\right) }T^{\ast }\left( x_{0}\right) .
\end{equation}%
It is clear that there is a peak of the Nernst signal at $x=10\%$ on the
positive side, in agreement with the Nernst effect experiments \cite%
{Wang1,Lee2,Wang2} in the cuprates. The signal weakens on either side of $%
x=10\%$ because the magnitude of the DDW order parameter weakens with $x$ on
either side of this value of doping.

\section{Conclusion}

In conclusion, we show that the Nernst signal from an ambipolar DDW state
has a robust low temperature peak which occurs below its mean field
transition temperature. The onset of the Nernst signal, however, occurs at
the transition temperature itself, which may account for the sizable Nernst
effect found in the experiments in the pseudogap phase of the high
temperature cuprate superconductors. The sign of the peak value of the
Nernst coefficient can be positive or negative depending on whether
electron- or hole-pockets, respectively, dominate the low temperature
thermoelectric transport. For the experimental situation in some cuprates,
where the low temperature Hall coefficient is found to be negative
indicating the dominance of the electron pocket in transport, we find that
the peak in the temperature dependent Nernst coefficient is on the positive
side. In contrast, there is no such peak when the DDW state is
non-ambipolar. In this case, with only one type of pockets in the excitation
spectrum, the Nernst signal is negative for both electron and hole pockets.
However, quite surprisingly, we find that these two individual negative
contributions `add up' to produce a net positive Nernst effect in the
ambipolar DDW state. We prove these results both by numerical calculations
as well as analytical arguments, which establishes the robustness of the
existence of the low temperature peak, making it insensitive to any
reasonable variations of the microscopic parameters.

With a reasonable assumption about the doping dependence of the DDW order
parameter, which is physically motivated and stipulates weakening of the
order parameter with hole doping on either side of the underdoped regime, we
find that the low temperature Nernst coefficient also has a pronounced peak
as a function of hole concentration. The peak of the Nernst coefficient
coincides with the value of doping where the DDW order parameter is assumed
to be the strongest, and the signal weakens on either side of this value of
the hole concentration. At low value of the hole doping, we find that the
Nernst coefficient remains negative over a wider range of temperature than
at moderate underdoping, $x=10\%$.

To derive these results, we model the pseudogap phase by a doping-dependent
DDW order parameter and assume transport scattering times which are constant
throughout the Fermi surface and also independent of energy in a small ($T$%
-dependent) energy interval around the Fermi energy. Even though our chosen
ambipolar DDW state (with its specific Fermi surface topology (Fig.~(\ref%
{fig:pockets})) is not the only possible state consistent with either the
recent quantum oscillation experiments ~\cite%
{Doiron-Leyraud:2007,LeBoeuf:2007,Bangura:2008,Jaudet:2008,Yelland:2008,Sebastian:2008}
or the observed negative Hall coefficient \cite{LeBoeuf:2007} in the
underdoped cuprates, it is at least one possible state qualitatively
consistent with both. As such, the calculations for the Nernst coefficient
in the ambipolar DDW state given in this paper present at least one possible
scenario for the observed enhanced Nernst signals in the underdoped cuprate
superconductors.

\textbf{Acknowledgement:} Zhang is supported by WSU start-up funds. S. T. is
supported by DOE/EPSCoR Grant No: DE-FG02-04ER-46139 and Clemson University
start up funds. S. C. is supported by Grant No: NSF DMR 0705092.


\begin{thebibliography}
\bibitem{} \expandafter\ifx\csname natexlab\endcsname\relax

\fi
\expandafter\ifx\csname bibnamefont\endcsname\relax

\fi
\expandafter\ifx\csname bibfnamefont\endcsname\relax

\fi
\expandafter\ifx\csname citenamefont\endcsname\relax

\fi
\expandafter\ifx\csname url\endcsname\relax

\fi
\expandafter\ifx\csname urlprefix\endcsname\relax

\fi
\providecommand{\bibinfo}[2]{#2} \providecommand{\eprint}[2][]{\url{#2}}

\bibitem{Norman} M. R. Norman, D. Pines, C. Kallin, Adv. Phys. \textbf{54},
715 (2005).

\bibitem{Halperin} S. Chakravarty, B. I. Halperin, and D. R. Nelson, Phys.
Rev. B \textbf{39}, 2344 (1989).

\bibitem{Taillefer} L. Taillefer, J. Phys.: Condens. Matter 21 164212 (2009).

\bibitem[Doiron-Leyraud et~al.(2007)Doiron-Leyraud, Proust, LeBoeuf,
Levallois, Bonnemaison, Liang, Bonn, Hardy, and Taillefer]%
{Doiron-Leyraud:2007} \bibinfo{author}{\bibfnamefont{N.}~%
\bibnamefont{Doiron-Leyraud}}, \bibinfo{author}{\bibfnamefont{C.}~%
\bibnamefont{Proust}}, \bibinfo{author}{\bibfnamefont{D.}~%
\bibnamefont{LeBoeuf}}, \bibinfo{author}{\bibfnamefont{J.}~%
\bibnamefont{Levallois}},
\bibinfo{author}{\bibfnamefont{J.-B.}
\bibnamefont{Bonnemaison}}, \bibinfo{author}{\bibfnamefont{R.}~%
\bibnamefont{Liang}},
\bibinfo{author}{\bibfnamefont{D.~A.}
\bibnamefont{Bonn}},
\bibinfo{author}{\bibfnamefont{W.~N.}
\bibnamefont{Hardy}}, and \bibinfo{author}{\bibfnamefont{L.}~%
\bibnamefont{Taillefer}}, \bibinfo{journal}{Nature} \textbf{%
\bibinfo{volume}{447}}, \bibinfo{pages}{565} (\bibinfo{year}{2007}).

\bibitem[LeBoeuf et~al.(2007)LeBoeuf, Doiron-Leyraud, Levallois, Daou,
Bonnemaison, Hussey, Balicas, Ramshaw, Liang, Bonn et~al.]{LeBoeuf:2007} %
\bibinfo{author}{\bibfnamefont{D.}~\bibnamefont{LeBoeuf }}, %
\bibinfo{author}{\bibfnamefont{N.}~\bibnamefont{Doiron-Leyraud}}, %
\bibinfo{author}{\bibfnamefont{J.}~\bibnamefont{Levallois}}, %
\bibinfo{author}{\bibfnamefont{R.}~\bibnamefont{Daou}}, \bibinfo{author}{%
\bibfnamefont{J.~B.} \bibnamefont{Bonnemaison}}, \bibinfo{author}{%
\bibfnamefont{N.~E.} \bibnamefont{Hussey}}, \bibinfo{author}{%
\bibfnamefont{L.}~\bibnamefont{Balicas}}, \bibinfo{author}{%
\bibfnamefont{B.~J.} \bibnamefont{Ramshaw}}, \bibinfo{author}{%
\bibfnamefont{R.}~\bibnamefont{Liang}}, \bibinfo{author}{%
\bibfnamefont{D.~A.} \bibnamefont{Bonn}}, et~al., \bibinfo{journal}{Nature}
\textbf{\bibinfo{volume}{450}}, \bibinfo{pages}{533} (\bibinfo{year}{2007}).

\bibitem[Bangura et~al.(2008)Bangura, Fletcher, Carrington, Levallois,
Nardone, Vignolle, Heard, Doiron-Leyraud, LeBoeuf, Taillefer et~al.]%
{Bangura:2008} \bibinfo{author}{\bibfnamefont{A.~F.} \bibnamefont{Bangura}}, %
\bibinfo{author}{\bibfnamefont{J.~D.} \bibnamefont{Fletcher}}, %
\bibinfo{author}{\bibfnamefont{A.}~\bibnamefont{Carrington}}, %
\bibinfo{author}{\bibfnamefont{J.}~\bibnamefont{Levallois}}, %
\bibinfo{author}{\bibfnamefont{M.}~\bibnamefont{Nardone}}, %
\bibinfo{author}{\bibfnamefont{B.}~\bibnamefont{Vignolle}}, %
\bibinfo{author}{\bibfnamefont{P.~J.} \bibnamefont{Heard}}, %
\bibinfo{author}{\bibfnamefont{N.}~\bibnamefont{Doiron-Leyraud}}, %
\bibinfo{author}{\bibfnamefont{D.}~\bibnamefont{LeBoeuf}}, %
\bibinfo{author}{\bibfnamefont{L.}~\bibnamefont{Taillefer}}, et~al., %
\bibinfo{journal}{Phys. Rev. Lett.} \textbf{\bibinfo{volume}{100}}, %
\bibinfo{pages}{047004} (\bibinfo{year}{2008}).

\bibitem[Jaudet et~al.(2008)Jaudet, Vignolles, Audouard, Levallois, LeBoeuf,
Doiron-Leyraud, Vignolle, Nardone, Zitouni, Liang et~al.]{Jaudet:2008} %
\bibinfo{author}{\bibfnamefont{C.}~\bibnamefont{Jaudet}}, %
\bibinfo{author}{\bibfnamefont{D.}~\bibnamefont{Vignolles}}, %
\bibinfo{author}{\bibfnamefont{A.}~\bibnamefont{Audouard}}, %
\bibinfo{author}{\bibfnamefont{J.}~\bibnamefont{Levallois}}, %
\bibinfo{author}{\bibfnamefont{D.}~\bibnamefont{LeBoeuf}}, %
\bibinfo{author}{\bibfnamefont{N.}~\bibnamefont{Doiron-Leyraud}}, %
\bibinfo{author}{\bibfnamefont{B.}~\bibnamefont{Vignolle}}, %
\bibinfo{author}{\bibfnamefont{M.}~\bibnamefont{Nardone}}, %
\bibinfo{author}{\bibfnamefont{A.}~\bibnamefont{Zitouni}}, %
\bibinfo{author}{\bibfnamefont{R.}~\bibnamefont{Liang}}, et~al., %
\bibinfo{journal}{Phys. Rev. Lett.} \textbf{\bibinfo{volume}{100}}, %
\bibinfo{pages}{187005} (\bibinfo{year}{2008}).

\bibitem[Yelland et~al.(2008)Yelland, Singleton, Mielke, Harrison,
Balakirev, Dabrowski, and Cooper]{Yelland:2008} \bibinfo{author}{%
\bibfnamefont{E.~A.} \bibnamefont{Yelland}}, \bibinfo{author}{%
\bibfnamefont{J.}~\bibnamefont{Singleton}}, \bibinfo{author}{%
\bibfnamefont{C.~H.} \bibnamefont{Mielke}}, \bibinfo{author}{%
\bibfnamefont{N.}~\bibnamefont{Harrison}}, \bibinfo{author}{%
\bibfnamefont{F.~F.} \bibnamefont{Balakirev}}, \bibinfo{author}{%
\bibfnamefont{B.}~\bibnamefont{Dabrowski}}, and
\bibinfo{author}{\bibfnamefont{J.~R.}
  \bibnamefont{Cooper}}, \bibinfo{journal}{Phys. Rev. Lett.} \textbf{%
\bibinfo{volume}{100}}, \bibinfo{pages}{047003} (\bibinfo{year}{2008}).

\bibitem[Sebastian et~al.(2008)Sebastian, Harrison, Palm, Murphy, Mielke,
Liang, Bonn, Hardy, and Lonzarich]{Sebastian:2008} \bibinfo{author}{%
\bibfnamefont{S.~E.} \bibnamefont{Sebastian}}, \bibinfo{author}{%
\bibfnamefont{N.}~\bibnamefont{Harrison}}, \bibinfo{author}{%
\bibfnamefont{E.}~\bibnamefont{Palm}},
\bibinfo{author}{\bibfnamefont{T.~P.}
\bibnamefont{Murphy}},
\bibinfo{author}{\bibfnamefont{C.~H.}
\bibnamefont{Mielke}}, \bibinfo{author}{\bibfnamefont{R.}~%
\bibnamefont{Liang}},
\bibinfo{author}{\bibfnamefont{D.~A.}
\bibnamefont{Bonn}},
\bibinfo{author}{\bibfnamefont{W.~N.}
\bibnamefont{Hardy}}, and
\bibinfo{author}{\bibfnamefont{G.~G.}
\bibnamefont{Lonzarich}}, \bibinfo{journal}{Nature} \textbf{%
\bibinfo{volume}{454}}, \bibinfo{pages}{200} (\bibinfo{year}{2008}).

\bibitem[Chakravarty and Kee(2008)]{Chakravarty:2008b} \bibinfo{author}{%
\bibfnamefont{S.}~\bibnamefont{Chakravarty}} and \bibinfo{author}{%
\bibfnamefont{H.-Y.} \bibnamefont{Kee}},
\bibinfo{journal}{Proc. Natl.
Acad. Sci. USA} \textbf{\bibinfo{volume}{105}}, \bibinfo{pages}{8835} (%
\bibinfo{year}{2008}).

\bibitem[Millis and Norman(2007)]{Millis:2007} \bibinfo{author}{%
\bibfnamefont{A.~J.} \bibnamefont{Millis}} and \bibinfo{author}{%
\bibfnamefont{M.~R.} \bibnamefont{Norman}}, \bibinfo{journal}{Phys. Rev. B}
\textbf{\bibinfo{volume}{76}}, \bibinfo{pages}{220503} (\bibinfo{year}{2007}%
).

\bibitem[Podolsky and Kee(2008)]{Podolsky:2008} \bibinfo{author}{%
\bibfnamefont{D.}~\bibnamefont{Podolsky}} and \bibinfo{author}{%
\bibfnamefont{H.-Y.} \bibnamefont{Kee}},
\bibinfo{journal}{Phys. Rev. B 78,
224516 (2008)} (\bibinfo{year}{2008}).

\bibitem{Morinari:2009} T. Morinari, J. Phys. Soc. Jpn, \textbf{78}, 054708
(2009).

\bibitem{Chen:2009} K-T. Chen and P. A. Lee, Phys. Rev. B \textbf{79},
180510 (R) (2009).

\bibitem{Subir} E. G. Moon, S. Sachdev, Phys. Rev. B \textbf{80}, 035117
(2009).

\bibitem{Shen} A. Damascelli, Z. Hussain, and Z.-X. Shen, Rev. Mod. Phys.
\textbf{75}, 473 (2003).

\bibitem{Jia} X. Jia, P. Goswami, and S. Chakravarty, Phys. Rev. B \textbf{80%
}, 134503 (2009).

\bibitem{Sudip-ARPES} S. Chakravarty, C. Nayak, S. Tewari, Phys. Rev B
\textbf{68}, 100504 (2003).

\bibitem{Meng} J. Meng, G. Liu, W. Zhang, L. Zhao, H. Liu, \textit{et. al.},
Nature \textbf{462}, 335 (2009).

\bibitem{Yang} H.-B. Yang, J. D. Rameau, P. D. Johnson, T. Valla, A.
Tsvelik, and G. D. Gu, Nature \textbf{456}, 77 (2008).

\bibitem{Wang1} Y. Wang, Z. A. Xu, T. Kakeshita, S. Uchida, S. Ono, Yoichi
Ando, and N. P. Ong, Phys. Rev. B. \textbf{64}, 224519 (2001).

\bibitem{Lee2} W.-L. Lee, S. Watauchi, V. L. Miller, R. J. Cava, and N. P.
Ong, Phys. Rev. Lett. \textbf{93}, 226601 (2004).

\bibitem{Wang2} Y. Wang, L. Li, and N. P. Ong, Phys. Rev. B, \textbf{73},
024510 (2006).

\bibitem{Olivier} O. Cyr-Choiniere, R. Daou, F. Laliberte, D. LeBoeuf, N.
Doiron-Leyraud, J. Chang, J.-Q. Yan, J.-G. Cheng, J.-S. Zhou, J.B.
Goodenough, S. Pyon, T. Takayama, H. Takagi, Y. Tanaka, L. Taillefer, Nature%
\textbf{\ 458}, 743 (2009).

\bibitem{chang} J. Chang, R. Daou, C. Proust, D. LeBoeuf, N. Doiron-Leyraud,
F. Laliberte, B. Pingault, B. J. Ramshaw, R. Liang, D. A. Bonn, W. N. Hardy,
H. Takagi, A. Antunes, I. Sheikin, K. Behnia, L. Taillefer, arXiv:0907.5039

\bibitem{Nayak} C. Nayak, Phys. Rev. B \textbf{62}, 4880 (2000).

\bibitem{Chakravarty01} S. Chakravarty, R. B. Laughlin, D. K. Morr, and C.
Nayak, Phys. Rev. B \textbf{63}, 094503 (2001).


\bibitem{Sumanta} S. Tewari, S. Chakravarty, J. O. Fjaerestad, C. Nayak, and
R. S. Thompson, Phys. Rev. B \textbf{70}, 014514 (2004).

\bibitem{Chakravarty04} S. Chakravarty, H.-Y. Kee, and K. Volker, Nature
\textbf{428}, 53 (2004).

\bibitem[Tewari et~al.(2001)Tewari, Kee, Nayak, and Chakravarty]%
{Tewari:2001} \bibinfo{author}{\bibfnamefont{S.}~\bibnamefont{Tewari}}, %
\bibinfo{author}{\bibfnamefont{H.-Y.} \bibnamefont{Kee}}, %
\bibinfo{author}{\bibfnamefont{C.}~\bibnamefont{Nayak}}, and %
\bibinfo{author}{\bibfnamefont{S.}~\bibnamefont{Chakravarty}}, %
\bibinfo{journal}{Phys. Rev. B} \textbf{\bibinfo{volume}{64}}, %
\bibinfo{pages}{224516} (\bibinfo{year}{2001}).

\bibitem{Tewari} S. Chakravarty, C. Nayak, S. Tewari, and X. Yang, Phys.
Rev. Lett.\textbf{\ 89}, 277003 (2002).

\bibitem[Nayak(2000)]{Nayak:2000} \bibinfo{author}{\bibfnamefont{C.}~%
\bibnamefont{Nayak}}, \bibinfo{journal}{Phys. Rev. B} \textbf{%
\bibinfo{volume}{62}}, \bibinfo{pages}{4880} (\bibinfo{year}{2000}).

\bibitem[Nayak and Pivovarov(2002)]{Nayak:2002} \bibinfo{author}{%
\bibfnamefont{C.}~\bibnamefont{Nayak}} and \bibinfo{author}{%
\bibfnamefont{E.}~\bibnamefont{Pivovarov}}, \bibinfo{journal}{Phys. Rev. B}
\textbf{\bibinfo{volume}{66}}, \bibinfo{pages}{064508} (\bibinfo{year}{2002}%
).

\bibitem{Mook1} H. A. Mook, P. Dai, S. M. Hayden, A. Hiess, J. W. Lynn, S.
H. Lee, and F. Do\u{g}an, Phys. Rev. B \textbf{66}, 144513 (2002).

\bibitem{Mook2} H. A. Mook, P. Dai, S. M. Hayden, A. Hiess, S. H. Lee, and
F. Do\u{g}an, Phys. Rev. B \textbf{69}, 134509 (2004).

\bibitem{Stock} C. Stock, W. J. L. Buyers, Z. Tun, R. Liang, D. Peets, D.
Bonn, W. N. Hardy, and L. Taillefer, Phys. Rev. B \textbf{66}, 024505 (2002).

\bibitem{Fauque} B. Fauqu\'{e}, Y. Sidis, V. Hinkov, S. Pailhes, C. T. Lin,
X. Chaud, and P. Bourges, Phys. Rev. Lett. \textbf{96}, 197001 (2006).

\bibitem[foo()]{footnote}
\bibinfo{note}{We note, however, that polarized neutron scattering and an analysis of the reciprocal space form factor, both crucial
 to uncovering the DDW order as emphasized in Refs.~\onlinecite{Mook1, Mook2}, were not performed in Ref.~\onlinecite{Stock}. Regarding Ref.~\onlinecite{Fauque}, we point out that no data proving the non-existence of DDW are
  presented.}

\bibitem{Hackl} A. Hackl, M. Vojta, S. Sachdev, arXiv:0908.1088.

\bibitem{Kivelson} S. A. Kivelson, I. P. Bindloss, E. Fradkin, V. Oganesyan,
J. M. Tranquada, A. Kapitulnik, and C. Howald, Rev. Mod. Phys. \textbf{75},
1201 (2003).

\bibitem{Anderson} O. K. Andersen, A. I. Liechtenstein, O. Jepsen, and F.
Paulsen, J. Phys. Chem. Solids \textbf{56}, 1573 (1995).

\bibitem{Tewari1} S. Tewari, and C. Zhang, Phys. Rev. Lett. \textbf{103},
077001 (2009).

\bibitem{Behnia} K. Behnia, J. Phys.: Condens. Matter \textbf{21}, 113101
(2009) .

\bibitem{Nolas} G. S. Nolas, J. Sharp and H. J. Goldsmid, \emph{%
Thermoelectrics}, Springer (2001).



\bibitem{Trugman} S. Trugman, Phys. Rev. Lett. \textbf{65}, 500 (1990).

\bibitem{Koralek} J. D. Koralek, J. F. Douglas, N. C. Plumb, Z. Sun, A. V.
Fedorov, M. M. Murnane, H. C. Kapteyn, S. T. Cundiff, Y. Aiura, K. Oka, H.
Eisaki, and D. S. Dessau, Phys. Rev. Lett. \textbf{96}, 017005 (2006).

\bibitem{Marder} N. W. Ashcroft and N. D. Mermin, \textit{Solid State Physics%
}, (Saunders College Publishing, New York, 1976).

\bibitem{Oganesyan} V. Oganesyan and I. Ussishkin, Phys. Rev. B \textbf{70},
054503 (2004).


\bibitem{Stephen} M. J. Stephen, Phys. Rev. B \textbf{45}, 5481 (1992).

\bibitem{Pallab} I. Dimov, P. Goswami, X. Jia, and S. Chakravarty, Phys.
Rev. B \textbf{78}, 134529 (2008).

\end{thebibliography}
\end{document}